# Enhancement of the Residual Linear Term in the Thermal Conductivity of Kondo Insulator SmB$_6$ under Uniaxial Tensile Strain

Brian Casas,[1*] Laura Yu,[1*] Zachary Fisk,[1] and Jing Xia[1]

[1]Department of Physics and Astronomy, University of California, Irvine, California 92697, USA
* equal contribution

Recent experiments on Kondo insulator SmB$_6$ have revealed bulk excitations that could be Fermions without charge. This would lead to a residual linear term in the thermal conductivity that was not observed universally in experiments. To solve this mystery, we introduce a novel symmetric geometry for performing thermal transport measurements under strains. Magnetic-field-independent residual linear terms were found in flux-grown SmB$_6$ at nominal zero strain, and they grew under a tensile strain, which is known to modify Sm valence. A small growth-dependent difference in the Sm valence might explain the observed discrepancies. We discuss the constraints on the theories of bulk excitations in SmB$_6$.

SmB$_6$ is a strongly correlated Kondo insulator where localized 4f electrons and the 5d conduction band interact to form a hybridization (Kondo) gap [1,2] that is responsible for the gradual insulating behavior in the bulk [1]. A resistance plateau appears at low temperatures below 5 $K$ due to the development of a robust [3] conducting surface [4,5], which is likely of a topological nature [3,6]. De Haas-van Alphen (dHvA) quantum oscillation [7] has revealed a 2D Fermi surface from the conducting surface state in flux-grown crystals, although caution needs to be taken with contributions from embedded epitaxial aluminum [8].

The bulk of SmB$_6$ is extremely insulating, with a divergence of the bulk DC resistivity over ten orders of magnitude towards the lowest temperatures [9]. Yet this simple picture of the bulk took an unexpected turn when another dHvA study performed on floating zone-grown crystals uncovered an unexpected 3D Fermi surface of the bulk [10,11]. To explain this finding, charge-neutral fermions have been theoretically proposed to exist in the insulating bulk of SmB$_6$ [10–12], and are responsible for the observed dHvA signal: a Fermi surface in the absence of a Fermi liquid. Anomalous excitations in the bulk have also been observed in THz transmission experiments on both flux and floating zone crystals [13], where high frequency (> 0.2 THz) bulk conductivity was found to be non-zero at the lowest temperature when the free carrier Drude contribution has already vanished completely [14].

To elucidate the nature of the mysterious bulk excitations, given the complication and discrepancies associated with dHvA experiments, temperature-dependent thermal conductivity $\kappa(T)$ measurements [11,15,16] have been carried out to probe the proposed bulk charge-neutral excitations that carry entropy. In principle $\kappa(T)$ could contain both Bosonic and Fermionic contributions $\kappa(T) = \kappa_{Bose}(T) + \kappa_{Fermi}(T)$. The former is dominated by phonons and takes the form $\kappa_{ph}(T) = b\,T^3$ at low enough temperatures [17]. In contrast, $\kappa_{Fermi}$ is temperature-linear ($\kappa_{Fermi} = a\,T$) at low temperatures [17], and could have contributions from both conducting Fermions $\kappa_{CF}(T)$ and insulating Fermions $\kappa_{IF}(T)$: $\kappa_{Fermi}(T) = \kappa_{CF}(T) + \kappa_{IF}(T)$, where $\kappa_{CF}$ can be estimated from the electric conductivity $\rho_0$ using the Wiedemann-Franz law as $\kappa_{CF}/T(T \to 0) \sim 10^{-4} - 10^{-6}\,WK^{-2}m^{-1}$ [15]. A "residual linear term" $a \equiv \kappa/T(T \to 0)$ far exceeding this estimated value would indicate a substantial $\kappa_{IF}$, and would serve as strong evidence for the proposed neutral Fermions.

Experimentally, however, $\kappa(T)$ measurements on crystals grown in different labs have yielded contradicting results: while $a$ on the order of $10^{-2}\,WK^{-2}m^{-1}$ have been reported in floating-zone samples [11], essentially zero values were found in flux-grown [16] and floating-zone [15] crystals that were grown elsewhere. Understanding and reconciling this profound discrepancy is a critical step towards testing the neutral Fermion proposal in SmB$_6$ and the understanding of the topological Kondo insulator.

It is well known that SmB$_6$ crystals, especially from the floating-zone method, could have non-stoichiometric crystal growth that results in Sm vacancies of around 1% concentration due to the vaporization of Sm during the growth process [18]. These vacancies cause an enhancement of the Sm valence from the nonmagnetic $Sm^{2+}$ to the $J = 5/2$ magnetic $Sm^{3+}$ [19]. Therefore, it has been hypothesized that a difference in Sm valence could induce a change in $a = \kappa/T(T \to 0)$ and account for the observed discrepancies. Since it is impractical to change the vacancy level after crystal growth, an alternative method is required to alter Sm valence and test this hypothesis. Sm ions in SmB$_6$ are known to be of mixed-valence, where recent studies have attributed the valence fluctuations to the $B_2$ dimer having a temperature and pressure-dependent stretch and displacement that causes a change in the electron configuration and charge transfer, affecting the valence of the Sm ions [20,21]. As a result, the Sm valence can be enhanced through mechanical pressure [22,23] or uniaxial compressive strain [24], providing a convenient platform for testing the above hypothesis. Here we report $\kappa(T)$ measurements of flux-grown SmB$_6$ under a varying uniaxial strain between 70 $mK$ and 12 $K$, using a novel symmetric thermal transport geometry on a uniaxial strain apparatus [24].

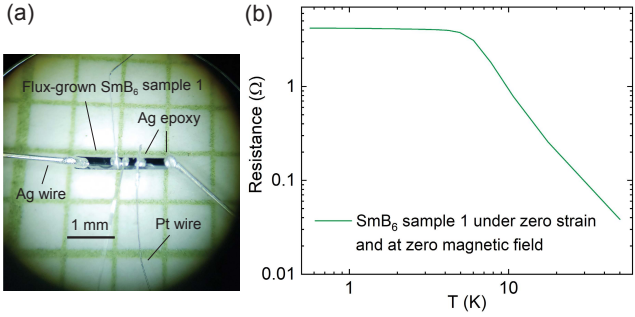

FIG. 1. (a) Flux-grown SmB$_6$ sample 1 attached with silver and platinum wires using silver epoxy for a thermal and electric transport measurements. (b) Unstrained resistance showing a plateau of 4 Ω. The fitted Kondo gap is 34.7 K.

SmB$_6$ crystals were grown using the flux method. As shown in Fig. 1(a), we chose crystals with a needle-shaped morphology with naturally grown <100> faces from the same growth batch. They are etched with 20% HCl solution for roughly 10 minutes to remove excess oxides and absorbed gases to ensure low contact resistance of less than 100 Ω. platinum wires 25 $\mu m$ in diameter are used to establish electric and thermal contacts via H20E silver epoxy cured at 80°C for three hours. As shown in Fig. 1(b), a resistance plateau of a very low saturation resistance of 4 Ohm occurs below 5 $K$, indicating high crystal qualities. Thermally activated behaviors were observed between 6 $K$ and 50 $K$ with a fitted activation gap of 34.7 $K$ that is consistent with crystals from previous growth batches [3,5,8].

Thermal conductivity measurements were taken in a dilution fridge and a He$^3$ fridge for different temperature ranges. As a check for phonon contribution $\kappa_{ph}(T)$, we first measure $\kappa(T)$ on sample 2 with the traditional steady-state method [10,12,13] in the high-temperature range when $\kappa_{ph}(T)$ dominates. There is a visible peak in $\kappa(T)$ near 9 $K$ as shown in Fig. 2(a), indicative of a transition between electron-phonon scattering and multi-phonon scattering channels. The peak height signifies high sample quality, as it is nearly 20% higher than the data previously reported [25]. Fig. 2(b) shows $\kappa(T)$ under various magnetic fields. Unlike the field-independent $\kappa(T)$ reported in [16], a non-monotonic magnetic field dependence is evident, which is similar to those reported in [15] due to the scattering of phonons by magnetic rare-earth impurities or vacancies.

Uniaxial strain is applied to the sample using a cryogenic tri-piezo strain apparatus that was first introduced by Hicks et. al. [26], and was later adopted for the electric transport studies of SmB$_6$ [24] as shown in Fig. 3(b). This apparatus allows a controllable application of strain with about 70% of the strain transferring from the piezo actuator to the SmB$_6$ sample, which was calibrated using a built-in strain gauge and optical imaging [24]. Opposite voltages are applied to middle and side piezo actuators to exert opposite displacements between middle and side titanium frames. During cooldown, the thermal contractions of the middle and side piezo actuators (and titanium frames) cancel and will not contribute to the residual strain. A residual compressive strain due to uncanceled differential thermal contraction between the SmB$_6$ crystal and the same length of titanium frame is estimated to be 0.1% and nearly temperature independent below a few Kelvins.

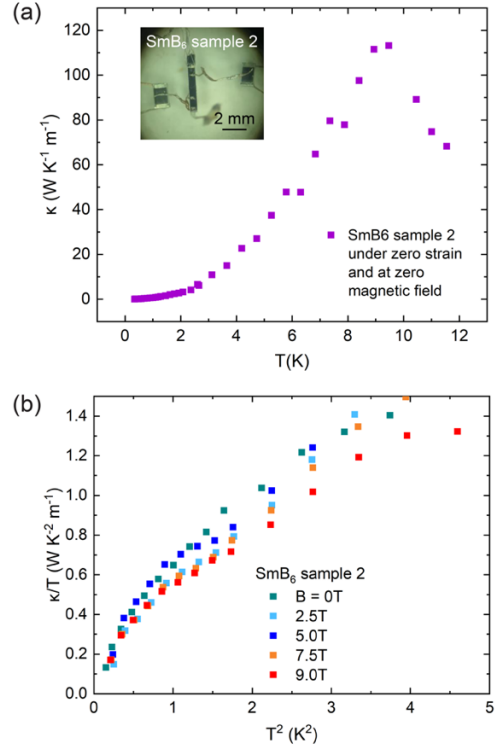

FIG. 2. $\kappa$ of unstrained sample 2 up to 12 $K$ where $\kappa_{ph}$ dominates. (a) the $\kappa(T)$ peak at 9 $K$ indicates high sample quality, (b) non-monotonic magnetic field dependence of $\kappa_{ph}$ due to the scattering of phonons by magnetic impurities.

Strong mechanical clamping in the form of Stycast 2850FT epoxy is required at both ends of the SmB$_6$ crystal mounted between the gap in the middle titanium frame. And this leads to a high degree of thermal contact at both ends of the sample that causes a severe challenge to the conventional steady-state thermal transport setup [10,12,13] where one end of the sample needs to be isolated from the thermal ground and connected to a thermally suspended heater.

To overcome this challenge, we have utilized a symmetric design and placed the heater at the center of the mounted crystal as shown in Fig. 3(a). Heat current is injected by the heater at the center and split off towards both ends. Two thermometers placed off-center measure the thermal gradient $\Delta T$ from which $\kappa(T)$ can be calculated. Considering the deviation of the heater's position from the center, the expression for thermal conductivity is $\kappa(T) = \frac{P}{\Delta T} \frac{l_R}{(l_L + l_R)} \frac{l}{A}$, where $P$ is the power dissipated from the heater, $\Delta T$ is the thermal gradient, $l_L$ and $l_R$ are distances between the heater and left and right clamping points, $l$ is the separation between the two thermometers, and $A$ is the cross-sectional area of the crystal. We note that the temperature dependence of $\kappa(T)$, which is the main subject of this paper, should remain accurate even in the presence of significant asymmetry or inaccuracy of these dimensional parameters, which are inevitable due to the complexity of the setup and the small size of the crystals.

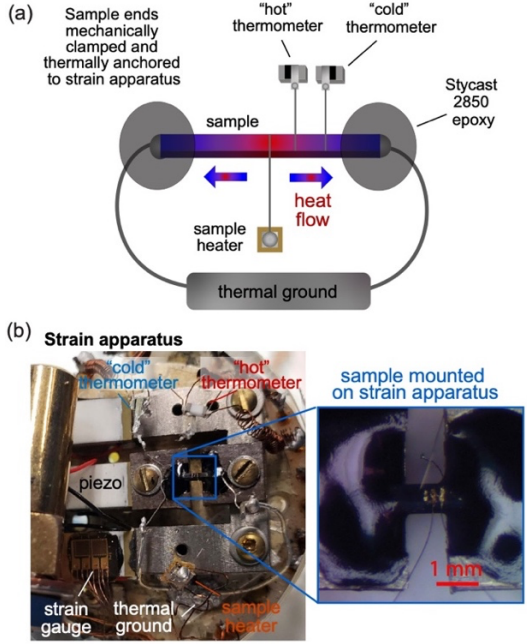

FIG. 3. (a) In our symmetric geometry, the sample is mechanically clamped and thermally grounded on both ends. Thermal gradient is established between the central heater, and both ends. (b) Photo of mounted sample 3.

Compared to the conventional measurement geometry in Fig. 2(a), the crystal under strain breaks easily during the clamping, cooling down, and straining, especially in the presence of imperfections and embedded aluminum flux. The latter, with a superconducting temperature of $\sim 1\,K$, has been screened with AC susceptibility. Being superconducting, trace amounts of aluminum will not contribute to $a = \kappa/T(T \to 0)$ at zero magnetic fields. In addition, the piezo actuators tend to fail after thermal cycling and voltage sweeping, which become a severe challenge during lengthy measurements limited by the long thermalization time at $mK$ temperatures. Nevertheless, we have obtained results on two crystals samples 1 and 3 with the strained symmetric thermal setup.

Sample 1 (cross-section $A = 0.241 \times 0.254\,mm^2$) was first measured with the conventional method [10,12,13]. It was then rewired and epoxy-mounted on the symmetric strain apparatus to repeat $\kappa(T)$ measurement. The results are plotted as $\kappa/T$ vs $T^2$ in blue and red in Fig. 4. Both data sets were fitted to the form $\kappa/T = a + bT^2$, where $a$ is the residual linear term and $b$ is the coefficient of phonon contribution. A theoretical phonon curve (green) is added as a guide to the eye, where $b_{Theory} = 0.455\,WK^{-3}m^{-1}$ is calculated using a phonon mean-free path that is roughly approximated by the square root of $A$ and other material parameters of $SmB_6$ [15]. The fitted parameter $b$ for phonon contribution $\kappa_{ph}(T) = b\,T^3$ is magnetic field dependent due to the scattering of phonons by magnetic impurities [15]. $b$ is also expected to be strain-dependent: compression increases $\kappa_{ph}(T)$ in ionic crystals [27]. Indeed, with the residual compressive strain ($\sim 0.1\%$) in the "symmetric setup" (blue in Fig. 4a), $b$ has increased by 5% from $0.306\,WK^{-3}m^{-1}$ to $0.321\,WK^{-3}m^{-1}$.

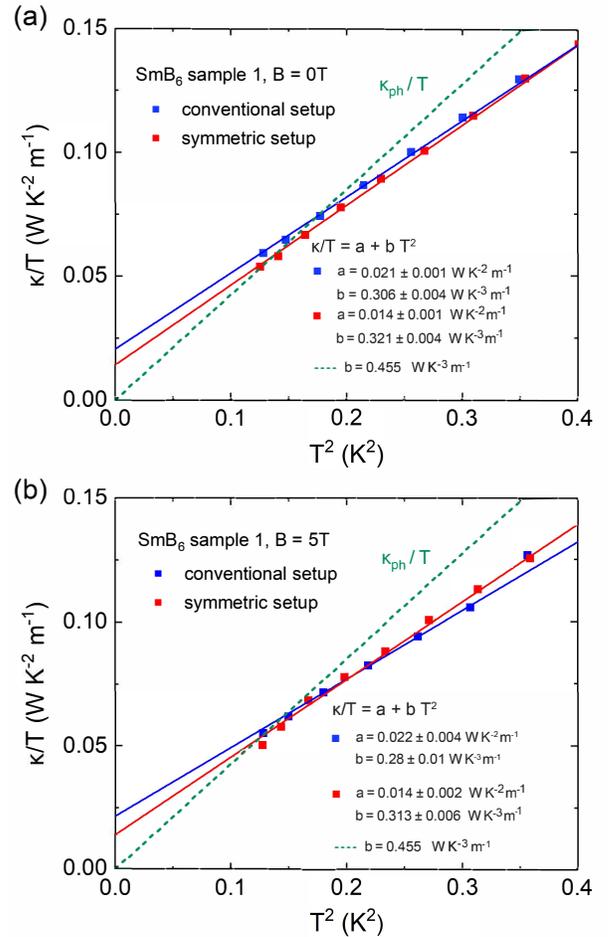

FIG. 4. $\kappa/T$ vs $T^2$ of sample 1 ($A = 0.241 \times 0.254$ mm$^2$) measured with conventional and symmetric setups: (a) under zero magnetic field and (b) under 5 $T$. The dashed lines show linear fits below 0.6 $K$. The difference in $\kappa(T)$ between the two methods is due to a residual compressive strain ($\sim 0.1\%$) in the symmetric strain setup during cooldown.

The most striking feature in Fig. 4 is the non-zero residual linear term that is *independent* of the applied magnetic field:

$a = 0.021(2)\ WK^{-2}m^{-1}$ with the conventional setup where the sample is suspended and truly unstrained, and is reduced by 30% to $a = 0.014\ WK^{-2}m^{-1}$ with the symmetric setup where a residual compressive strain of 0.1% is present. These $a$ values are similar to those reported in the floating zone samples [11]. The large 30% reduction in $a$ is too large to be ascribed to geometric uncertainties during the crystal remounting to the strain setup and thus must stem from the 0.1% residual compressive strain.

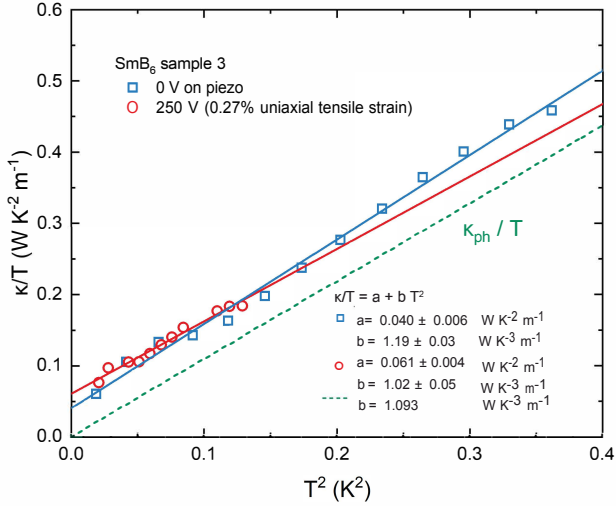

FIG. 5. $\kappa/T$ vs $T^2$ of SmB$_6$ sample 3 (A = 0.635 × 0.558 mm$^2$) under nominally zero strain and 0.27% tensile strain.

A definitive evaluation of the strain dependence of the residual linear term $a$ requires the modulation of applied strain *in situ*. Sample 1 broke in this process. And sample 3 (cross-section $A = 0.635 \times 0.558$ mm$^2$) was mounted on the strain setup with zero piezo voltage during the cooldown to 50 *mK*. And it was slowly warmed up with 250 V piezo voltage until it broke at 400 *mK*. The measured thermal conductivities (blue and red) are plotted in Fig. 5, along with a theoretical phonon curve (green) estimated from its cross-section 0.635 × 0.558 mm$^2$. Like in sample 1, the phonon contribution is strain-dependent: upon the application of 250 V to the piezo actuators, corresponding to an increase of tensile strain of 0.27%, $b$ is reduced by 14%.

At the same time, this increase of the tensile strain has dramatically enhanced the residual linear term $a$ by 50% from $0.040\ WK^{-2}m^{-1}$ to $0.061\ WK^{-2}m^{-1}$. As $a$ must be non-negative, the trends observed in samples 1 and 3 suggest that the residual linear term $a$ could remain at a minimum value of zero with a larger (> 0.3%) compressive strain, although we haven't been successful in applying a large negative piezo voltage without breaking the sample. A 0.3% compressive strain is roughly equivalent to 1 $GPa$ of pressure in SmB$_6$ [24], which would increase Sm valence by only 0.02 [23]. This indicates that the observed discrepancies [11,15,16] in the residual linear term may be explained by assuming a tiny growth-dependent difference in the Sm valance, which could arise due to concentration variations of Sm vacancies or residual strains introduced during high-temperature growth.

Now we discuss the nature of the Fermions that give rise to the observed strain-*dependent* and magnetic-field-*independent* linear term $\kappa_F(T) = a\,T$. Quite recently, one-dimensional dislocations have been revealed by chemical etching imaging [9] in the bulk of SmB$_6$, showing a density of surface termination points of $10^5\ cm^{-2}$ in floating zone samples and $2 \times 10^3\ cm^{-2}$ in flux samples. The metallic surfaces of these 1D "strips" could in principle contribute to both electrical and thermal conduction. However, the carriers are charged and would thus follow the Wiedemann-Franz law. And an upper bound of $a$ could be estimated as $a \sim L_0 l / AR_0 \sim 2 \times 10^{-5}\ WK^{-2}m^{-1}$, where $L_0 = 2.44 \times 10^{-8}\ W\Omega K^{-2}$, cross-sectional area $A \sim 3 \times 10^{-7}\ m^2$, crystal length $l \sim 10^{-3}\ m$, and resistance plateau $R_0 \sim 4$ Ω. Therefore, these 1D dislocations alone are not enough to account for the observed $a \sim 10^{-2}\ WK^{-2}m^{-1}$. For the same reason, our observations can't be explained by the nodal semimetal proposals [28,29]. It seems that the Wiedemann-Franz law must break down in the bulk. One scenario is assuming charge-neutral Fermions, whose density and/or scattering rate need to couple to strain, but not to the magnetic field. Charge-neutral fermions have indeed been proposed in the bulk of SmB$_6$ [10–12], but were primarily discussed to explain the bulk dHvA signal. We note that such bulk dHvA signals have not been observed in our flux crystals except if embedded epitaxial aluminum is present [8]. Therefore, a modified neutral Fermion model would be necessary to account for all the experimental observations.

In summary, we have observed a strain-*dependent* and magnetic-field-*independent* linear term $\kappa_F(T) = a\,T$ in the bulk of SmB$_6$, which necessitates bulk neutral Fermions and puts constraints on theories. The introduced symmetric geometry enables future *mK* thermal transport studies of crystals under a tunable strain.

This work is supported by NSF awards DMR-1807817 and DMR- 1708199. J.X. has been supported in part by the Gordon and Betty Moore Foundation through Grant GBMF10276. J.X. conceived the experiments and designed the symmetric thermal geometry. B.C. and L.Y. carried out the measurements, fabricated the samples, and performed data analysis. J.X. and Z.F. supervised the project. We acknowledge assistance and discussion with L. Li, L. Chen, V. Galitski, S. Thomas and P. Rosa during the early stage of this project.

Email: xia.jing@uci.edu